\documentclass[lettersize,journal]{IEEEtran}
\usepackage{amsmath,amsfonts}
\usepackage{algorithmic}
\usepackage{algorithm}
\usepackage{array}
\usepackage[caption=false,font=normalsize,labelfont=sf,textfont=sf]{subfig}
\usepackage{textcomp}
\usepackage{stfloats}
\usepackage{verbatim}
\usepackage{graphicx}
\usepackage{cite}
\usepackage{CJKutf8}
\usepackage{multirow}
\usepackage{soul}
\usepackage[normalem]{ulem}
\usepackage{xspace}
\usepackage{graphicx}
\usepackage{xcolor}
\usepackage{enumitem}
\usepackage{titlesec}
\usepackage{url}

\definecolor{gold}{RGB}{218,165,32}
\definecolor{purple}{RGB}{160, 32, 240}

\newcommand{\revision}[1]{\leavevmode{\color{blue}{#1}}}
\newcommand{\remark}[1]{\textcolor{red}{#1}}
\newcommand{\removed}[1]{\leavevmode{\color{red}{\sout{#1}}}}
\newcommand{\mylabel}[1]{\textbf{#1}}

\def \cleanversion{}
\ifx\cleanversion\undefined
\else
 \renewcommand{\remark}[1]{} %
 \renewcommand{\mylabel}[1]{} %
 \renewcommand{\removed}[1]{} %
 \renewcommand{\revision}[1]{#1}
\fi

\def\promptvis{PrompTHis\xspace}

\def\ivg{Image Variant Graph\xspace}

\renewcommand{\thesubsubsection}{\arabic{subsubsection})}
\titleformat{\subsubsection}[block]{\normalfont\normalsize\itshape}{\thesubsubsection}{1em}{}
\titlespacing*{\subsubsection}{0em}{0em}{0em}

\newcommand{\paratitle}[1]{{\sffamily #1}}

\usepackage{graphicx}
\usepackage{scalerel}

\begin{document}

\title{\promptvis{}: Visualizing the Process and Influence of Prompt Editing during Text-to-Image Creation}

\author{Yuhan Guo, Hanning Shao, Can Liu, Kai Xu, and Xiaoru Yuan
\thanks{Yuhan Guo, Hanning Shao, Can Liu, and Xiaoru Yuan are with Key Laboratory of Machine Perception (Ministry of Education), School of AI, Peking University, China. E-mail: \{yuhan.guo, hanning.shao, can.liu, xiaoru.yuan\}@pku.edu.cn. Xiaoru Yuan is also with National Engineering Laboratory for Big Data Analysis and Application, Peking University. Xiaoru Yuan is the corresponding author.}
\thanks{Kai Xu is with University of Nottingham, UK. E-mail: Kai.Xu@nottingham.ac.uk.}
\thanks{Manuscript received xx; revised xx.}}

\markboth{Journal of \LaTeX\ Class Files,~Vol.~xx, No.~xx, xx}%
{Shell \MakeLowercase{\textit{et al.}}: A Sample Article Using IEEEtran.cls for IEEE Journals}

\maketitle

\begin{abstract}
Generative text-to-image models, which allow users to create appealing images through a text prompt, have seen a dramatic increase in popularity in recent years.
However, most users have a limited understanding of how such models work and it often requires many trials and errors to achieve satisfactory results.
The prompt history contains a wealth of information that could provide users with insights into what have been explored and how the prompt changes impact the output image, yet little research attention has been paid to the visual analysis of such process to support users.
We propose the \textit{\ivg}, a novel visual representation designed to support comparing prompt-image pairs and exploring the editing history.
The \ivg models prompt differences as edges between corresponding images and presents the distances between images through projection.
Based on the graph, we developed the \promptvis system through co-design with
artists. 
Besides \ivg, \promptvis also incorporates a detailed prompt-image history and a navigation mini-map.
\revision{Based on the review and analysis of the prompting history, users can better understand the impact of prompt changes and have a more effective control of image generation.}
\revision{A quantitative user study with eleven amateur participants and qualitative interviews with five professionals and one amateur user were conducted to evaluate the effectiveness of \promptvis. The results demonstrate \promptvis can help users review the prompt history, make sense of the model, and plan their creative process.}
\end{abstract}

\begin{IEEEkeywords}
Text visualization, image visualization, text-to-image generation, editing history, provenance, generative art
\end{IEEEkeywords}

\section{Introduction}

In recent years, generative text-to-image models, such as Stable Diffusion~\cite{stable-diffusion} and DALL-E 2~\cite{dalle2}, have gained significant popularity. %
These models can generate exquisite images from a text prompt, reducing the barriers for the general public to engage in visual creation and providing new avenues for artistic expression.
Many artists also begun to explore creative ideas with such models, taking advantage of the sometimes unexpected results they produce. 

Despite the enormous potential, it is often challenging to generate images that match artists' intentions and creative preferences. Most users have a limited understanding of such models and struggle to convey their intentions in a way that the model can understand. There is no guarantee of satisfactory outcomes even after many trials and errors. 
This is further complicated by the inherent randomness such models have, e.g., same prompt can lead to different images in different runs, and the fact that the mapping from language to images is ambiguous such that small change in the prompt can lead to large change in the resulting image. 
Together with the lack of support for organizing and reviewing previous attempts, 
artists often engage in near-random explorations, easily lose track of previous attempts, which often lead to repetitive efforts, stuck in the local convergence, or spending a significant amount of time without achieving desired outcomes.

In this work, we aim to address this challenge by designing a novel visual analytics system that can help artists better make sense of the behavior and characteristics of the generative model, which can in turn lead to a more efficient and effective creative process. We co-designed with artists who utilize generative models as part of their creative process, understand their goals, the current practice and workflow, and the challenges they face.
Two of the main needs we identified are that artists would like to know the image space that has been already explored to avoid repetition and understand how the changes in prompts influence the generation of images.

The term ``prompt engineering'' has been created to describe the methods and processes that help users create effective prompts.
Some research efforts have been devoted to creating visualization tools that can assist users in prompt editing or recommendation~\cite{wang2023reprompt, Feng2023promptmagician, Brade2023promptify}. However, these mostly try to match user prompts with previous examples, ignoring potential differences in individual's intention and preference. 
We took a different approach and focus on the prompt editing process, believing the prompt history contains the information that is key to a solution.
Before the prevalence of text-to-image models, the editing process primarily refers to revising textual content~\cite{Viegas2004historyflow, Chevalier2010animated, Guo2023edit}.
The arrival of the generative models has changed the nature of such editing, as two modalities, text and images, are involved.
Understanding these two modalities simultaneously poses great challenges.
Moreover, the two are connected, i.e., changes in the prompt text cause updates in the resulting images, and such connections are often complex and difficult to understand, if possible at all.

As a result, we developed the \textit{\ivg}, which models the prompt history as a graph with the images as nodes and the differences in text prompts as edges.
We assign weights to the edges to reflect how the modifications of a specific word impact the generation of images.
The \ivg positions the image nodes in a 2D space based on their visual similarity.
This allows users to observe the distribution of generated images and help analyze the impact of prompt change on the generation.

Based on \ivg{}, \promptvis{} is an interactive visualization system designed to further assist artists with prompt engineering. 
With the \ivg as the main view, the \promptvis{} system also includes a detailed prompt-image pair history, a prompt mini-map for navigation, \revision{and a creation panel to generate images.} %
A formal user study with eleven participants was conducted to evaluate effectiveness of \ivg and \promptvis in a post-analysis setting.
\revision{We further conducted in-depth interviews with five professional users and one amateur user to understand how the system supports the creative process. All participants found the system helpful for reviewing and understanding the prompt history. Three of the interview participants were engaged in leveraging \ivg to refine prompts according to model behavior and plan the creative process.}
To summarize, the contribution of our method can be summarized as follows:

\begin{enumerate} %
\item \ivg, a novel and efficient visual design for prompt history that reveals the image distribution from the existing attempts and how word-level changes in prompt influence the generation of images.
\item \promptvis, A visual analysis system that helps users explore the prompting history and make sense of the generative model through analysis of the editing history.
\item \revision{A user study and in-depth interviews to demonstrate the effectiveness of \ivg and \promptvis.}
\end{enumerate}

\section{Related Work}
\label{sec: back}

This section begins with the related literature in the field of text-to-image generation.
This is followed by the recent work on prompt engineering, particularly the support for the prompt editing process, because our work targets the process by individual creators to iteratively refine prompts to create artistic paintings.
The last part covers the related work in the broader field of visualization for the editing process.

\vspace{-0.1cm}
\subsection{Text-to-Image Generation}

Generative AI has attracted a huge amount of interest from the general public and professionals, since it demonstrated ground breaking capability in image creation.
Ever since OpenAI releases CLIP~\cite{radford2021learning} in DALL-E architecture~\cite{Ramesh2021Zero}, a contrastive language-image pre-training model that aligns natural languages and images in vector-based representations, a number of models are proposed to generate images from text, including VQGAN-CLIP~\cite{Crowson2022VQGAN} and latent diffusion~\cite{stable-diffusion}.
Please refer to a recent survey~\cite{survey-text2image} for more details.

These text-to-image models significantly reduce the barriers of creating images.
Artists are also very interested in these AI generators, not necessarily using the output as their work but more of an inspiration for creative ideas. The randomness and uncertainty during the generation process may lead to surprising results, some of which the artists we worked with found inspirational.
In addition, such models allow artists to quickly test out different ideas.
Therefore, many artists include these models as part of their creative workflow.

One of the main challenges faced by the artists when working with the generative models is how to compose effective text prompts, i.e. how to construct descriptions that can accurately capture their intention and preferences and also be understandable by the model.
How to create effective prompts becomes a craft itself, which is referred to as \textit{prompt engineering}.
Oppenlaender~\cite{oppenlaender2022taxonomy} summarized six prompt modifiers applied by individuals in the online community.
Liu et al.~\cite{liu2022design} also conducted experiments to analyze the influences of prompt keywords and model hyper-parameters on the outputs.
These works provide experimental guidelines for prompt constructions to help individuals produce better. However, these guidelines are usually model-specific, i.e., they do not always apply to different models. Also, each artist would have his or her own style and preference, and these nuanced differences are not always captured by these guidelines.

\vspace{-0.1cm}
\subsection{Auto and Visual Assistance in Prompt Engineering}
\label{sec:related_recommend}

Given the challenges of creating effective prompts, research has been carried out to help with prompt engineering.
In the context of text-to-text generation, PromptAid~\cite{mishra2023promptaid} helps users apply perturbations on keywords, paraphrases, and in-context examples to test and refine their prompts.
PromptIDE~\cite{strobelt2022interactive} allows prompt testing on small datasets before being applied to the whole dataset.
There are also a few methods designed specifically for text-to-image creation.
Some works visualize the details during the process of generation~\cite{lee2023diffusion} and allow user users to assign different prompts on different areas on the canvas and stages of the generation process~\cite{chung2023promptpaint}.
While useful, these methods are more suitable for users with enough technical knowledge and can benefit from the appreciation of internal process of generative models, which is often not the case for the members of the creative community.

Other works target non-technical users.
Wang et al.~\cite{wang2023reprompt} explored the emotional expressiveness in prompts from real users and proposed the RePrompt model to automatically refine users' prompts with emotion descriptions.
Promptify~\cite{Brade2023promptify} leverages large language models to recommend prompts.
PromptMagician~\cite{Feng2023promptmagician} extracts
similar prompt-image pairs from the DiffusionDB~\cite{wang2022diffusiondb} according to users' inputs and provides multi-view interaction that can help users find interesting recommendations and refine their prompts.

Shared among these methods is the approach to provide recommendations based on the similarity between user prompt and previous examples stored in a large database. However, as mentioned earlier, each artist may have his or her own style and preference, and previous examples may not be a good fit just because the prompts are similar.
Our work focuses on the analysis of the prompt engineering process. There are two potential benefits of this approach: first, it provides users with a more intuitive understanding of how the prompt changes impact the output images, allowing better control of the generation process without exposing the internal workings of the models. 
Second, it provides a more nuanced understanding of user intention and preference, which can be used to improve similarity-based recommendations. %

\vspace{-0.1cm}
\subsection{Visualization for Editing Process}

Our work focuses on the prompt editing history, and an important aspect of that is the changes of the prompt text. There are previous visualization methods designed for text comparison, though not in the context of text-to-image generation. Some of these methods focus on the comparison between different versions of the text, which is also known as 
``parallel texts''.
One of the common technique for visualizing parallel texts is juxtaposition~\cite{yousef2020survey, Viegas2004historyflow, janicke2017interactive}, often leveraging close and distant reading methods~\cite{moretti2005graphs, janicke2015close, alharbi2020transvis}.
The other important aspect of prompt history is the images generated at each step.
Research efforts have been made to visualize the changes in a collection of images.
These methods often employ projection methods to map the images to a 2D or 3D space to help reveal the similarities and differences among the images~\cite{Bertucci2022, Xie2019, Yang2006, janecek2003searching}.

In text-to-image generation, prompts and images are tightly coupled in the editing process and need to be considered together. Thus, existing visual comparison methods for text or images discussed earlier are not easily applicable.
In our work, prompts and corresponding images are always considered as pairs.
The \ivg visualizes the changes in both the prompt text and resulting images.
Together with the other features, \promptvis allows users to gain a better understanding of the relationships between the two
in a way that is different from other attempts so far.

The text and images involved in the prompt history can be considered as part of the \textit{Analytic Provenance}~\cite{analytic-provenance} of the creative process. Analytic provenance includes a wide range of contextual information about the analysis~\cite{xu2020survey}, from the data used, %
user interactions (which include the prompt), the analysis performed (such as the running of the generative model), intermediate and final results (e.g. images generated), to the user's critical thinking and analytic reasoning.
One of the common goals of analytic provenance analysis is to identify patterns in user behavior and reveal user's intention and analysis strategies.
From this perspective, \promptvis aims to better understand user's creative intention through the collection and analysis of the prompt history. %
While both intention and prompt/image are part of the analytic provenance, the former is much harder to capture and has to be inferred from the latter~\cite{provenance-agenda}.
However, if solved, even in a specific application context such as text-to-image generation, this would enable many exciting features, such as more effective recommendation and adaptive system~\cite{provenance-personalisation}.
This is the long-term goal of \promptvis: currently it focuses on the collection and visualization of analytic provenance; if successful the results would allow future work to provide more intelligent and nuanced support for artistic creations with generative models.

\section{Design Rationale}
\label{sec: space}

To understand how artists utilize text-to-image models in their creative workflow and their needs during this process, we conducted in-depth interview with two artists, including observation of their current practice. We started from learning about the artists themselves, such as their technical background and creative interests. We then went through their current workflow and observed a few examples. Finally we discussed with the artists about their experiences, comments, and challenges about the generative AI. 
The interview was recorded and thematic analysis was applied to the interview transcript. We summarize the interview and analysis results below.

\subsection{Workflow}

One of the artists experimented with the Disco-Diffusion via Colab notebook~\footnote{Disco Diffusion (Colab), accessed Sep 2023. Available at: \url{https://colab.research.google.com/github/alembics/disco-diffusion/blob/main/Disco_Diffusion.ipynb}.}.
and the other artist used Stable Diffusion~\cite{stable-diffusion} and Midjourney~\cite{midjourney}.
The artists often do not have a specific idea to start with and would adapt the prompt and setting as they progress.
The iterations stop when the artist is satisfied with the results, does not know how to further improve the prompt, or simply runs out of time. The latter two are far more common than the first one.
\revision{We organize the experiences and needs of the artists into the following insights.

\begin{itemize}[leftmargin=*]
    \item \textit{I1. Lack of organization in exploration process.} Currently there is no easy way to save the prompt/setting history and the generated images \revision{in Colab notebook}, so the artist manually created files and folders to save and organize them. Even though some current apps support saving the attempts automatically, it is not easy to review the explored settings and the extent to which the outcomes match expectations. As a result, the artists sometimes make repetitive unsatisfactory experiments. In other times they might temporarily leave an intermediate result and explore other branches, but forget or find it challenging to come back.
    \item \textit{I2. Misalignment between user intention and model output.} The model does not always understand user's intention in the prompt. In some cases, the model misinterpret the context of the prompt due to ambiguity in natural language. For example, once there were word ``head'' and ``shoulder'' in the prompt and shampoo appeared in the image. This was not intended and the artist would modify the prompt to remove the shampoo. But generating unexpected output is not always bad. Another time, to the artist's surprise, the output image had robotic animals that the artist liked and he then added the term to the next prompt.
    \item \textit{I3. Diverse requirements for personalized recommendations.} The artists put forward various desires for the model to make suggestions. Possible aspects of the recommendation include automatic exploration of parameters, guidance for refining prompts, and assessment of output images according to user's taste. The common emphasis is that the suggestions must cater to the preference of the artist.
\end{itemize}}

\subsection{Requirements}
\label{sec:requirements}

\revision{We are aware that direct guidance and recommendations (\textit{I3}), if effective, will significantly empower generative art creation. However, these require a thorough understanding of the users' requirements and preferences, which are reflected in the exploration history. 
Therefore, we choose to focus on the prompting history first and use the results as the foundation for more active support in the next step. The \textbf{target user} of this paper is professional artists who utilize generative models to explore creative ideas. The \textbf{usage scenario} is twofold: 1) reviewing and planning the creative process (\textit{I1}), and 2) making sense of the model's behavior so as to convey user's intention in a way that the model can understand (\textit{I2}). We believe such support serves as the foundation for understanding user intention and preference, which will enable us to create personalized recommendations (\textit{I3}). Therefore, we have summarized the following design requirements.}

\begin{itemize}[leftmargin=*]

\item \textbf{R1. Support organization and review of previous prompts and images.}
As is described in \textit{I1}, the artists spent a large amount of time in almost random trials and errors. %
Even after the prompts and images are saved, there is no easy way to review previous attempts and understand what has worked and what has not. 
There is a need to significantly reduce or eliminate the overhead of saving prompt history and a more effective means to organize and review previous attempts, even when the number of attempts is large.

\item \textbf{R2. Support comparison between different prompts and images.}
Text-to-image models like Disco Diffusion are guided by CLIP models~\cite{radford2021learning}, which aligns texts and images.
The artists find it helpful to ``understand how CLIP works'' by experimenting with different wording and comparing the results, so that they can set more accurate prompts and convey the intent to the model.
Currently, it is difficult to locate relevant text and images and compare them.

\item \textbf{\revision{R3. Provide users with a better understanding of model behavior.} }
While the ability to compare individual prompt and image would help artists understand the model's behavior at a micro level, there is also a need to understand such behaviors at a macro level, such as comparing two groups of prompts.
This would remove the negative impact the inherent randomness that generative models have,
and would also allow artist generalize their understanding of the underlying model, e.g., which types of prompts work well.

\item \textbf{\revision{R4. Help users plan the creative exploration process.} }
We observed several instances where artists consistently obtain unsatisfactory results regardless of adjustments made.
While any \textit{\revision{direct} support} (such as recommendations) would deserve a separate paper, we believe there is also the possibility of \textit{\revision{indirect} support}, such as  \revision{helping artists build and maintain a mental map of and orient themselves in the spaces explored,} which would help identify the gaps not covered so far and allow them to conduct creative explorations more systematically.

\end{itemize}

\vspace{-0.1cm}
\subsection{Overall Design}
For a creative session, while prompts and images are explicitly connected through the prompt-image pairs,  their distribution in the respective text and image space can be very different.
It is difficult to mentally align the semantic distribution of prompts and that of the images they generate, and we believe this is one of the main causes of the difficulties artists face.
For \promptvis, we chose to base the visualization on the image space and include information from the text space, since the artists' goal is the image and not the prompt.

The prompt history can be considered from two perspectives.
The first is the temporal evolution that represents artist's creative process.
The second is the semantic relations among various prompt-image pairs.
Given that one of the main goals is to help artists better understand the behaviors of the generative model (\textbf{R2}, \textbf{R3}), we decided to emphasize semantic relationships among prompts and images, e.g., how they are different or similar. The \promptvis system centers around the \ivg that models the differences in prompts as a variant graph that is common in text analysis.
The nodes are images and the edges are word-level differences in prompts. Therefore, users can explore how the text modifications affect the image generation (\textbf{R2}). \ivg also provides an overview of all the attempts (\textbf{R3}) and allows easy inspection of each of them through interaction (\textbf{R1}). Finally, the overview also allows identification of gaps in the exploration, \revision{helping user plan the creative process} (\textbf{R4}).
The details of the \ivg are discussed in Section~\ref{sec:graph}, whereas additional features in \promptvis{} are covered in Section~\ref{sec:system}.

\begin{figure}[htbp]
    \centering
    \vspace{-0.3cm}
    \includegraphics[width=1\linewidth]{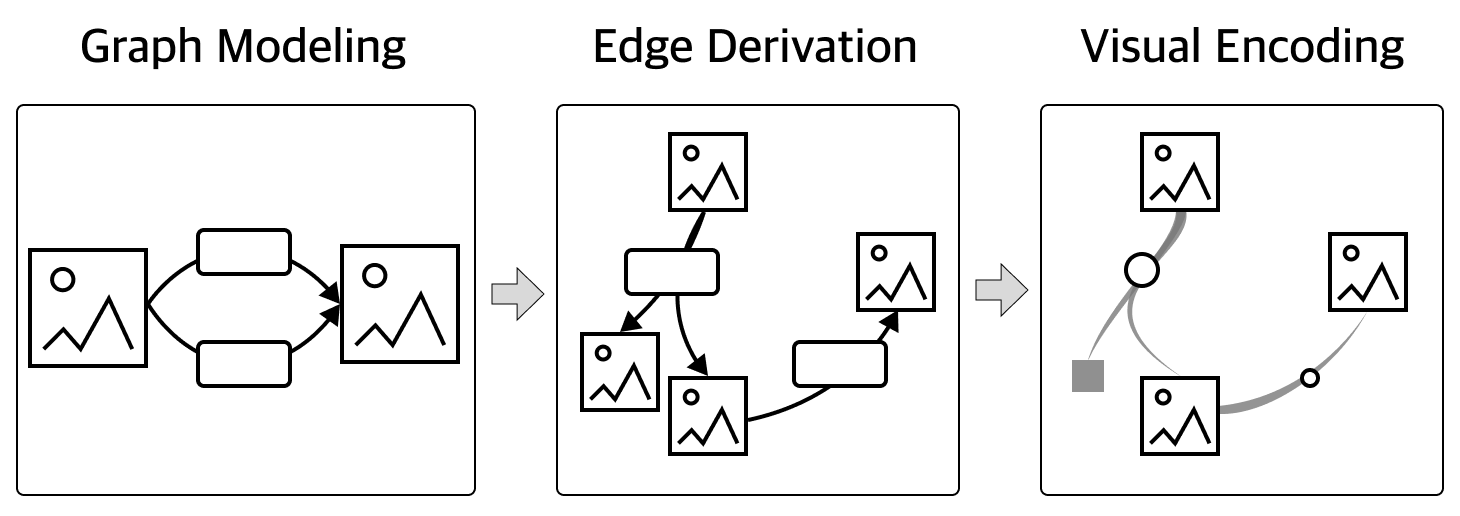}
    \vspace{-0.7cm}
    \caption{In \ivg the nodes are the images and the edges are the difference between prompts, one edge for each difference. Weighting algorithms are then applied to filter out less important edges. Finally, a novel layout algorithm and visual encoding are used to enhance scalability.}
    \label{fig:graph_construct}
    \vspace{-0.5cm}
\end{figure}

\section{Image Variant Graph}
\label{sec:graph}

\ivg aims to enable better understanding of the behaviors of text-to-image models through efficient comparison between text-image pairs (\textbf{R2}, \textbf{R3}) and allow easier navigation of large prompting histories (\textbf{R1}).
Fig.~\ref{fig:graph_construct} shows the conceptual construction pipeline of the \ivg, which is explained in more detail in the following sections.

\vspace{-0.1cm}
\subsection{Graph Modeling}
\label{sec:graph_model}

An important and challenging aspect that artists are concerned with is how modifications to prompts affect the generation of images.
As is shown in Fig.~\ref{fig:graph_construct}, each image is a node.
The word-level differences in prompts between two images are modeled as multiple edges connecting the two nodes, with each edge representing one-word insertion or deletion.

\begin{figure}[htbp]
    \centering
    \vspace{-0.3cm}
    \includegraphics[width=1\linewidth]{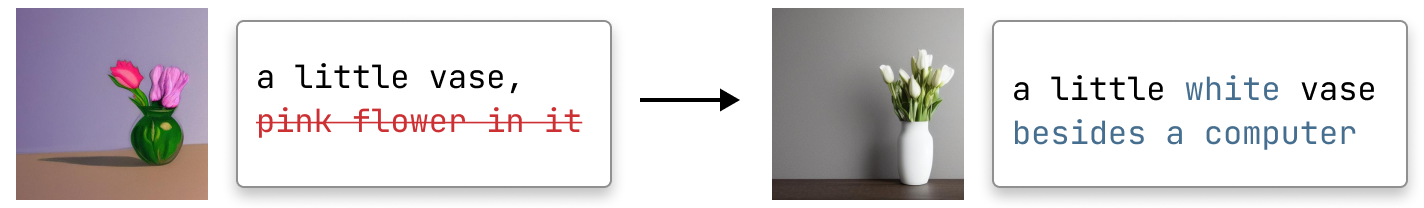}
    \vspace{-0.7cm}
    \caption{An example showing that not all the word modifications have the same impact on the image: While ``white'' causes the color of the vase to change, ``besides a computer'' does not have an obvious impact.}
    \label{fig:word_effect_example}
    \vspace{-0.25cm}
\end{figure}

As discussed earlier, \ivg emphasizes the semantics difference between prompts and not their temporal orders.
As a result, the \ivg is a complete graph, but in practice, not all the edges are important for explaining the image differences.
When the difference between two prompts involves multiple words, the impact of these words on the generated images can vary.
For instance, in Fig.~\ref{fig:word_effect_example}, the word ``white'' has a more prominent impact on the new image than the phrase ``besides a computer''.
A negative impact of showing all possible edges is that it can cause significant visual clutter. 
Therefore, we designed an algorithm to measure the edge \textit{weights}, i.e., the significance of the influence of edges, and use it to filter out less important edges. The details of the algorithm and its usage in the layout are discussed in Section~\ref{sec:edge} and Section~\ref{sec:graph_layout} respectively.

We considered a few alternatives when designing the graph model.
The first choice is whether to represent the text difference and the image variation in separate views or couple them in a single view.
We chose the latter due to the difficulty in aligning multiple text-image pairs in separate views.
The next design choice is how to show all the relevant information: 
text distribution, image distribution, text variations, image variations, relations between text and image distribution, and relations between text and image variations.
We chose to focus on the differences, as these are the most important aspects for users to understand the prompt change impact (\textbf{R2}, \textbf{R3}), using position for image difference and edges for text difference. This design also provides a good representation of the other four types of information (text/image distribution and relationship between distribution/variation).

\begin{figure}[htbp]
    \centering
    \vspace{-0.3cm}
    \includegraphics[width=1\linewidth]{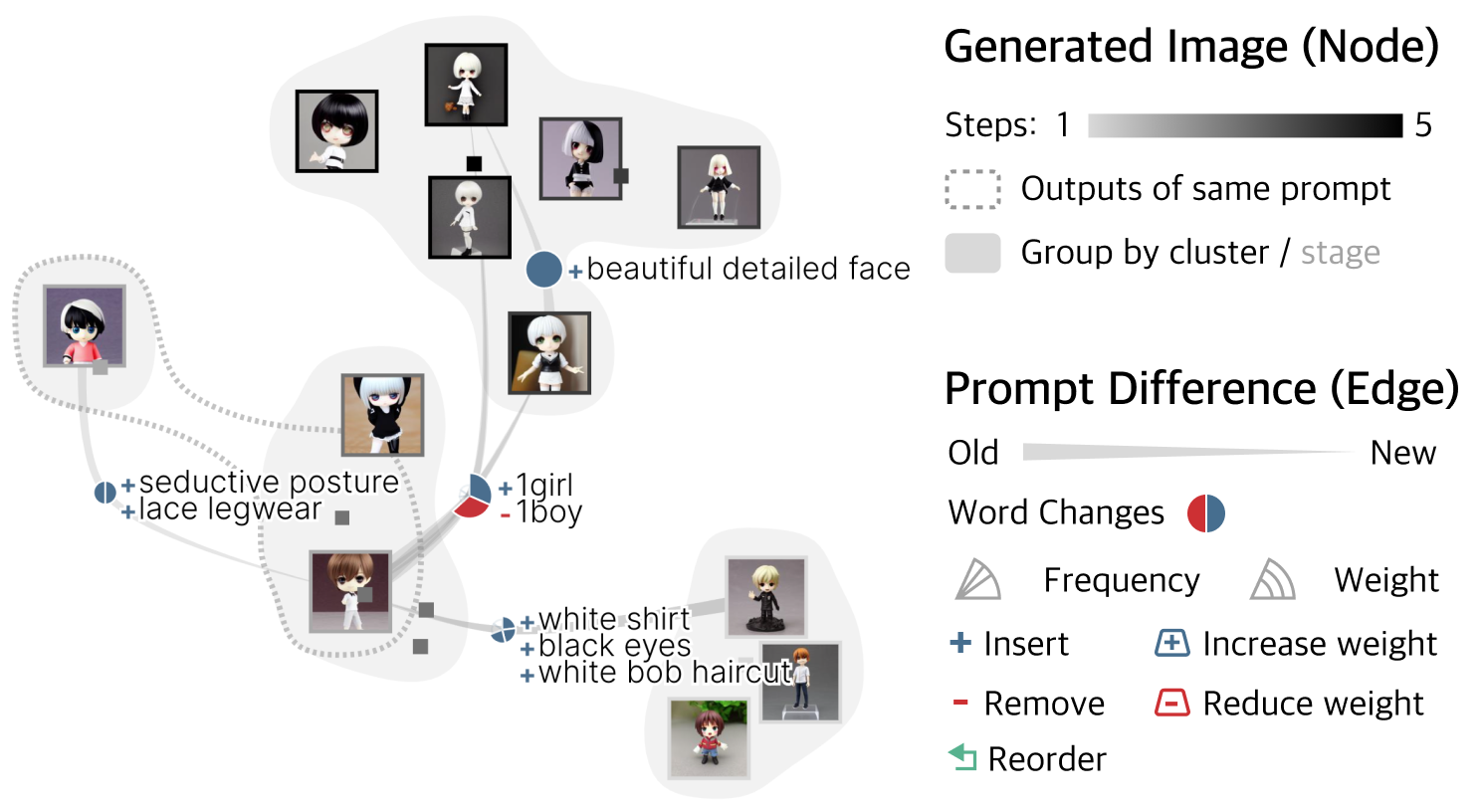}
    \vspace{-0.7cm}
    \caption{Visual encoding of \ivg. \revision{Image relationships are indicated by bubbles} and the word modifications are represented by glyphs.}
    \label{fig:graph_encoding}
    \vspace{-0.5cm}
\end{figure}

\vspace{-0.3cm}
\subsection{Visual Encoding}
\label{sec:graph_encoding}
As is illustrated in Fig.~\ref{fig:graph_encoding}, images are shown as thumbnails whose positions indicate the image variation. 
To reduce visual clutter, edges with the same word changes are bundled together. Less important images are represented by rectangles.

\textbf{Image nodes.}
Each image is represented as a thumbnail or small rectangle (if overlapped with more important images) scaled proportionally to the original one.
The gray scale of the rectangle encodes the temporal order, and so does the border of the shown image \revision{(top right of Fig.~\ref{fig:graph_encoding})}. %
When using text-to-image models, a single prompt often generates a batch of images, and the images within the same batch can vary significantly.
In \ivg, the image locations reflect similarity, based on which the images are clustered
(more on this in Section~\ref{sec:edge}). As a result, images from the same prompt could be far apart in the \ivg.
\revision{If the outputs of the same prompt fall into different clusters, a bubble with dashed border is added (e.g., bottom left of Fig.~\ref{fig:graph_encoding}).}
\revision{Bubbles with fill are used to enhance the visual representation of images within the same group, i.e., in the same cluster (Fig.~\ref{fig:graph_encoding}) or exploration stage (Fig.~\ref{fig:interface}).}

\textbf{Bundled edges.}
\revision{Text modifications are encoded as tapered edges, which is shown to be the most effective visual representation of edge direction~\cite{holten2009user}.}
Edges share the same text modifications or the same sources and targets are bundled together.
The actual word changes between the source and target images are represented by a glyph before the edge label.
\revision{Since there might be multiple words changed between the sources and targets, and only changes with higher \textit{weight} (please see Section~\ref{sec:edge} for details) are shown, the glyph encodes the change of each word as a slice of the circle.}
As is shown in the bottom right of Fig.~\ref{fig:graph_encoding},
\revision{the angle of the slice represents the frequency of the change among all the word modifications on the bundled edges, and the radius represents the weight. Slices considered as less important (with lower weight) are presented in low opacity.}
\textit{\textcolor[HTML]{466E8F}{Blue}} represents addition, either \textit{inserting} a new word or the \textit{increased weight} of a word.
\textit{\textcolor[HTML]{CD3033}{Red}} represents the subtraction, either the \textit{removal} of a word or a \textit{reduction in weight}.
\textit{\textcolor[HTML]{57B28F}{Green}} encodes \textit{reorder}.
\revision{For example, in Fig.~\ref{fig:graph_encoding}, the glyph, annotated with ``+1girl'' and ``-1boy'', indicates that the major cause of the image variation from the middle cluster to the top cluster is changing ``1boy'' to ``1girl''.}

\begin{figure*}[tb]
    \centering
    \vspace{-0.2cm}
    \includegraphics[width=1\linewidth]{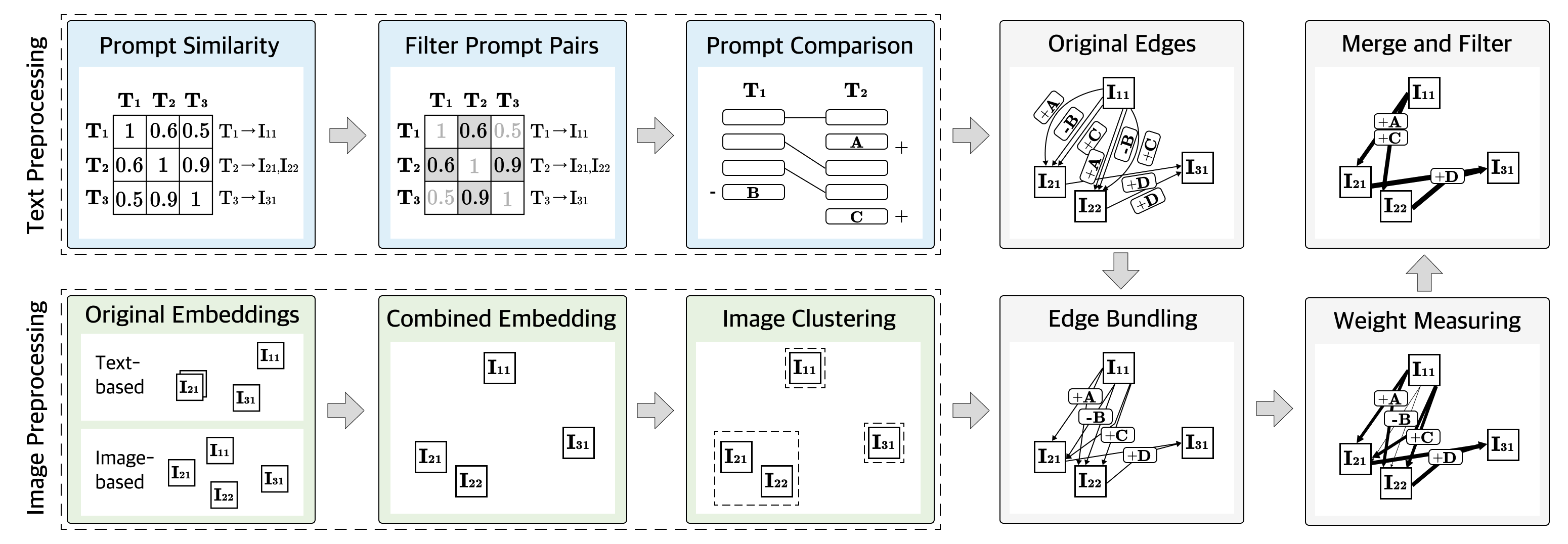}
    \vspace{-0.75cm}
    \caption{Pipeline of edge derivation. The text pre-processing stage compares the prompts to identify the word modifications and derive the original set of edges. Image pre-processing involves \revision{embedding images based on text and image encoding, combining the embedding,}
    and clustering images. Edges are then bundled based on the clusters. The impact of word modification on image change is calculated as edge weight, which is used to filter out low-impact edges.}
    \label{fig:edge_derive}
    \vspace{-0.5cm}
\end{figure*}

\vspace{-0.25cm}
\subsection{Edge Derivation}
\label{sec:edge}

The workflow of deriving edges is shown in Fig.~\ref{fig:edge_derive}.
We first compare the prompts and embed and cluster the images.
The original set of edges is obtained by comparing prompts and then bundled based on the image clusters.
After that, we calculate the edge \textit{weight}, which reflects the amount of image update the text change causes.
Based on the weights, the edges are further merged and filtered for visualization.

\begin{figure}[htbp]
    \centering
    \includegraphics[width=1\linewidth]
    {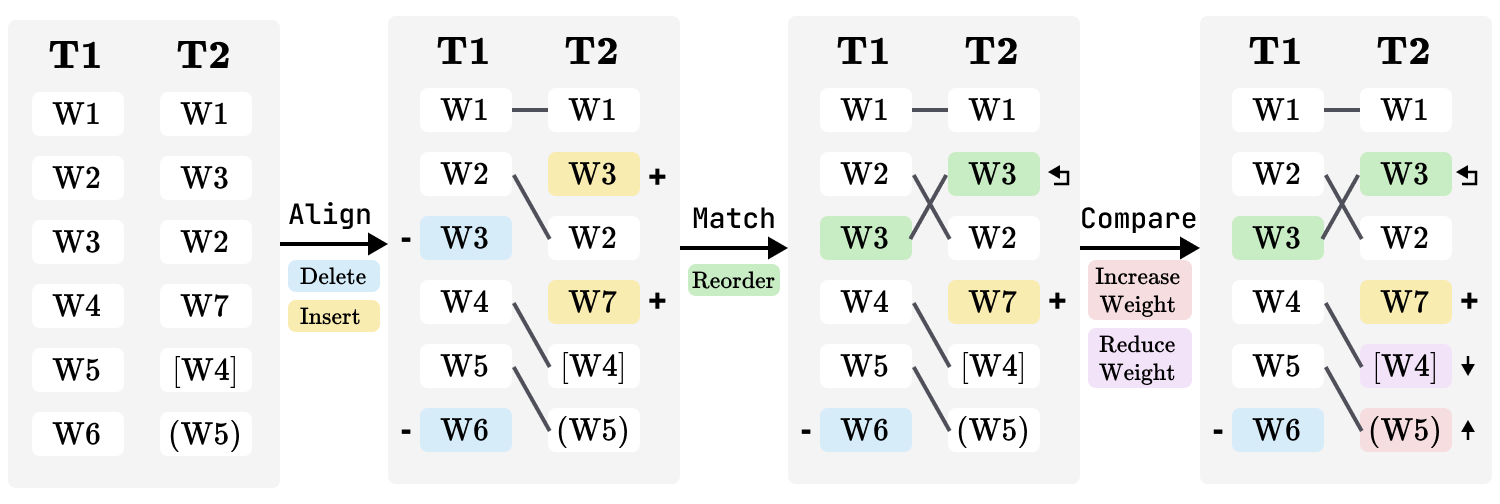}
    \vspace{-0.8cm}
    \caption{Three-step comparison of two prompts to identify word-level modifications. First, the Myers comparison algorithm~\cite{myers1986ano} is applied to calculate the inserted and deleted words. Then, the changed words are matched to identify the reordered words. Finally, the weights of the matched words are compared.}
    \label{fig:text_cmp}
    \vspace{-0.25cm}
\end{figure}

\textbf{Text pre-processing.}
The first step of text pre-processing is to calculate the \revision{Jaccard similarity} between every pair of prompts, resulting in a distance matrix.
We treat phrases the same way as words, i.e., if several words always appear together in the prompts, they are treated as one.
Only prompt pairs that are relatively similar are compared.
Prompt pairs with a distance \revision{higher} than the predefined \revision{lower} bound \revision{$S_{min}$} are reserved.
By default, \revision{$S_{min}$} is set as \revision{0.6}, \revision{and users can adjust the threshold to include more edges when the prompts vary significantly, or exclude edges if most prompts are similar}.
For each pair of prompts, we split the prompts into \textit{words} and compare the words to identify the modifications.
Diffusion models allow users to set the weight of specific words or phrases in a prompt following the given syntax.
The weights are parsed when splitting the prompts and each word is assigned a weight value (1 by default).
Fig.~\ref{fig:text_cmp} illustrates the comparison algorithm.
First, the Myers algorithm~\cite{myers1986ano} is applied to align the words, which identifies the \textit{insert} and \textit{remove} modification.
If a word is identified as removed in the first prompt and inserted in the second prompt, it is considered as \textit{reordered}.
Finally, the weights of the aligned words are compared to identify the \textit{increase weight} and \textit{reduce weight} operation.
Therefore, the edge can be denoted as $e = (w, a, I_{src}, I_{tgt})$, where $w$ denotes the modified word, $a$  denotes the modification action, $I_{src}$ and $I_{tgt}$ denotes the source image node and the target image node.

\textbf{Image pre-processing.}
In order to showcase the differences between images interpreted by the generative model, \revision{images are embedded to the two-dimensional space and grouped into clusters. We take both text information and image information into consideration for the embedding. Images are first encoded by the text encoder and image encoder of CLIP~\cite{radford2021learning} respectively. Each encoder transform the images into 512-dimensional vectors. The vectors are reduced to two dimensions through the t-SNE algorithm~\cite{van2008visualizing} with the cosine distance as the metric parameter, resulting in two groups of image embeddings, one based on the text space and the other based on the image space. The two spaces are aligned using Procrustes analysis and combined to generate the final embeddings. By default, the combined embedding is the average of the two, and users can adjust the weight of combination.}
Based on the embeddings, the images are clustered using the hierarchical agglomerative clustering algorithm.

\textbf{Edge bundling.}
The original edges are bundled according to the results of the image clustering,
since images within the same cluster tend to share more common features than images between clusters. The edges representing the same word modification with the starting node and ending node in the same cluster are bundled together.
Formally, two edges are bundled together if and only if $w$, $a$, $C(I_{src})$, and $C(I_{tgt})$ are the same for them, where $C(I)$ denotes the ID of the cluster which image $I$ belongs to.
Thus, the bundled edge can be denoted as $E = (w, a, C_{src}, C_{tgt})$.

\textbf{Weight measuring.}
\textit{Weight} is designed to quantify the impact a text modification has on the image change.
First, the more the changed words between two images are, the smaller the average impact of each word change would be.
Second, between two clusters of images, edges that better align with the common differences in prompts between the two groups are more likely to cause image variations.
That is, for example, if word A appears frequently in prompts of cluster one and is not included in cluster two, the modification of removing word A, which is represented by an edge from cluster one to cluster two, probably contributes to the difference.
We assume that the sum of the edge weights between two images is always one.
Initially, for every prompt pair, the edges between them are assigned equal weights.
Specifically, if there are $n_1$ images associated with the prompt $T_1$, $n_2$ images associated with the prompt $T_2$, the weight of each edge between these two image groups is set as $1 / (n_1 \cdot n_2 \cdot m)$, where $m$ is the distance between $T_1$ and $T_2$.
The distance $m$ is the number of different words which is calculated during the text preprocessing stage and illustrated in Fig.~\ref{fig:text_cmp}.
The weight of the bundled edge is the sum of the weights of its child edges.
$$W(E) = \sum\nolimits_{e \in E} W(e),$$
where $W(E)$ denotes the weight of the bundled edge $E$ and $W(e)$ denotes the weight of a child edge $e$.
However, not all edges between two images contribute equally to the image variation (Fig.~\ref{fig:word_effect_example}).
Based on the weights of the bundled edges, we redistribute the weights of each individual edge.
$$W(e) = \frac{W(E)}{\sum_{e' = (w', a', I_{src}, I_{tgt})} W(E')},$$
where $e'$ denotes any edge between $I_{src}$ and $I_{tgt}$, $E$ ($E'$) denotes the bundled edge that $e$ ($e'$) belongs to.
The weights of the bundled edges are updated accordingly.

\textbf{Merging and filtering.}
When the weights are updated, some multi-edges located between two images may still have the same weight, indicating that the algorithm cannot distinguish the difference in their impact on the image through the prompt history.
To reduce the abundance, we merge the multiple edges with the same weight into a single edge.
For the merged edge, there will be multiple word modifications and a glyph summarizes the edits (Section~\ref{sec:graph_encoding}).
The bundled edges whose weight is lower than the threshold $W_{min}$ will not be rendered without user demand.
\revision{By default, $W_{min}$ is calculated subject to the constraint that there are at most $N_E$ (we set $N_E$ as 12) edge bundles, and users can adjust the value $W_{min}$ to show fewer or more bundled edges.}

\vspace{-0.1cm}
\subsection{Layout and Drawing}
\label{sec:graph_layout}

In \ivg the nodes are positioned according to the embedding project and words are positioned at the barycenter of the source and target nodes.
To reduce clutter, we only show the thumbnail of representative images and present the rest of the images as glyphs.

\textbf{Image rendering.}
The image nodes are positioned according to the two-dimensional embeddings obtained during the image preprocessing stage shown in Fig.~\ref{fig:edge_derive}.
We measure the weights of the nodes according to the weights of the edges, i.e., the weight of a node is the sum of the weights of edges that starts from or ends at it.
Images are sorted in descending order based on their weights.
Each time the image with the largest weight is added if it does not overlap with any existing node. %
\revision{The bubbles are drawn with bubble sets~\cite{collins2009bubble}.}

\textbf{Word glyph positioning.}
Each bundled edge group has a glyph indicating the changed words.
This glyph also serves as the bunding point of the edges, which is positioned at the barycenter of the sources and targets of the edges.

\begin{figure}[htbp]
    \centering
    
    \vspace{-0.3cm}
    \includegraphics[width=1\linewidth]{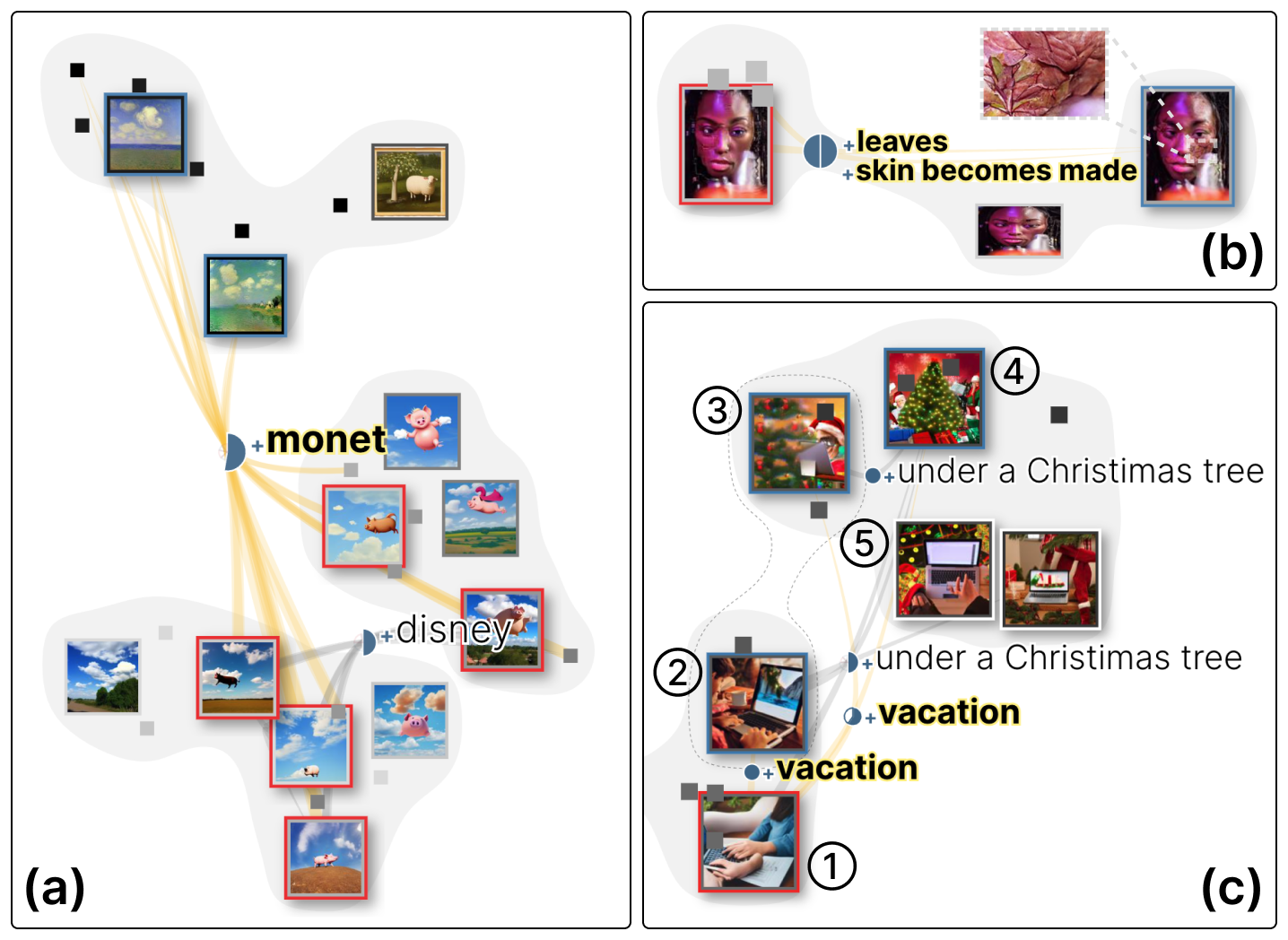}
    \vspace{-0.7cm}
    \caption{Different patterns of the influence of word modifications on the model's generation. (a) dominance (b) fine-tuning (c) association.}
    \label{fig:ivg_result}
    \vspace{-0.6cm}
\end{figure}

\vspace{-0.1cm}
\subsection{\revision{Results}}

\revision{Through node embedding and edge bundling, \ivg reveals how the word modifications influence the generation. Fig.~\ref{fig:ivg_result} shows three examples of the impact patterns.

``Dominance'' refers to the case where a slight change in the prompt causes a significant variation in the style or content of the images. For example, among the three clusters in Fig.~\ref{fig:ivg_result}a, the top one is generated by prompts such as \textit{``a pig in the sky, in \textcolor[HTML]{466E8F}{monet} style''}, the middle by \textit{``a pig in the sky, in disney style''}, and the bottom by \textit{``a pig in the sky''}, \textit{``a pig in the sunny and blue sky''}, etc. The edges from the bottom and middle clusters converge into the top cluster, indicating the word modification, inserting \textit{``\textcolor[HTML]{466E8F}{monet}''} dominates the style of the outcome. The dominant word can even take over the major character ``pig'' in the generation. For example, the top left image is an impressionist painting and there is no pig in the scene. At other times, newly added words introduce additional features to the previous prompt without destructing the original semantic, allowing for fine-tuning to the image. As shown in Fig.~\ref{fig:ivg_result}b, the left image is generated by prompt \textit{``a black woman is taken over by robotic flesh, 80s computer graphics overlay her face,''} and a detailed description \textit{``skin becomes made of leaves''} is added to change the skin texture (the right image).

Not all words have a stable and expected impact on the generated results. One type of these words is concept that does not describe a certain object but can evoke associations. As shown in Fig.~\ref{fig:ivg_result}c, edges representing inserting \textit{``\textcolor[HTML]{466E8F}{vacation}''} are bundled into two branches, one pointing to the same cluster and the other pointing to the top cluster presenting a Christmas tree. Specifically, c1 is generated by \textit{``playing computer in holiday''}, c2 and c3 are generated by \textit{``playing computer in holiday, vacation''}, and c4 and c5 by \textit{``playing computer in holiday, vacation, under a Christimas tree''} (there is a typo in the prompt, but the model recognizes the intended word ``Christmas''). Although ``vacation'' is a relatively abstract concept, it often co-occurs with ``Christmas'' in real-world data. This may be the reason why the model associates the word with a Christmas tree.}

\begin{figure*}[htbp]
    \centering
    \includegraphics[width=1\linewidth]{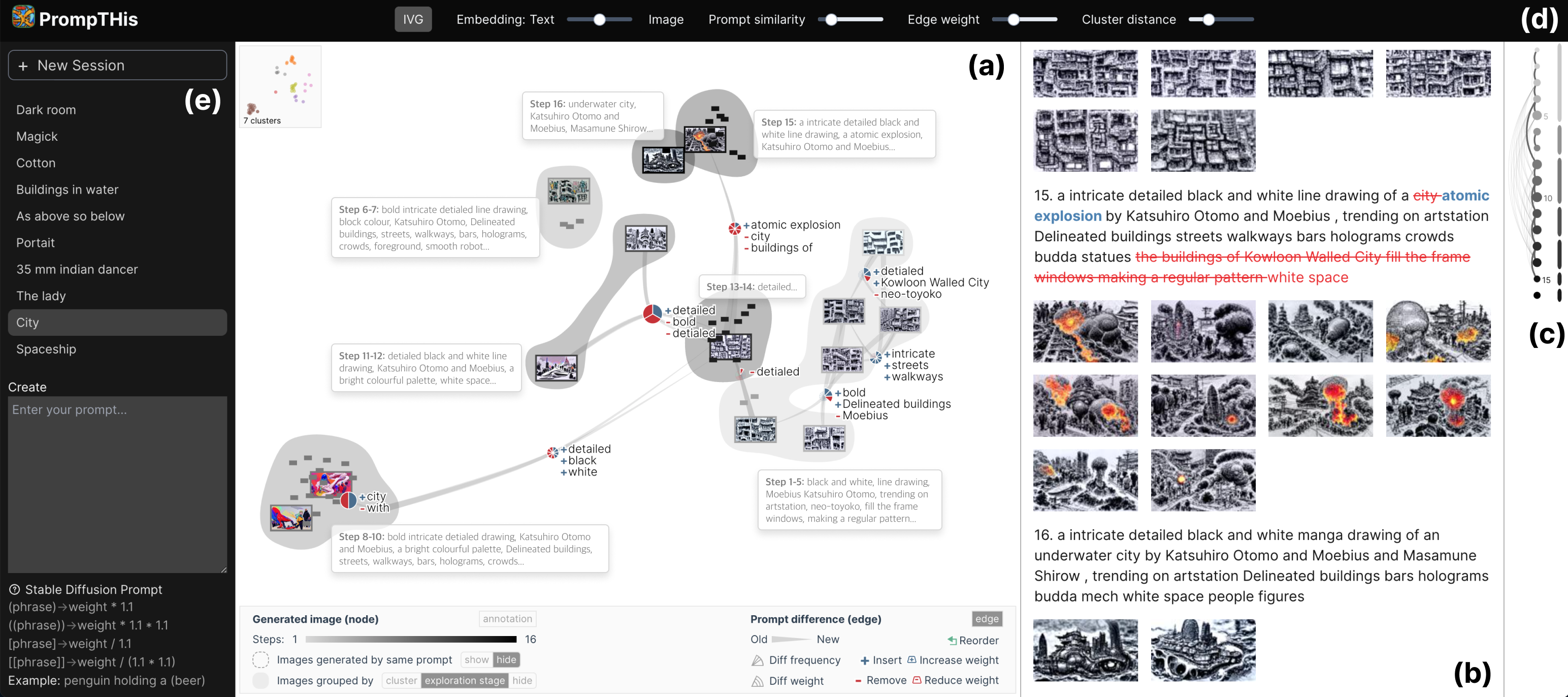}
    \vspace{-0.7cm}
    \caption{\revision{The interface of \promptvis, which consists of \ivg~(a), a history box~(b), a navigation mini-map~(c), a control panel~(d), and a creation panel~(e).}
    This figure shows the prompting records of an artist.
    Starting from a black-and-white drawing of city buildings (1-5), the artist experimented with color styles (6-7, 8-10), and returned to the black-and-white style (11-14), with ``atomic explosion'' inserted later (15).}
    \vspace{-0.55cm}
    \label{fig:interface}
\end{figure*}

\section{\promptvis System}
\label{sec:system}

\promptvis{} is a prototype designed to support artists to understand, navigate, and manage the prompting history during their creative workflow with generative text-to-image models.
As is shown in Fig.~\ref{fig:interface}, \ivg is the main view of the system, which supports users to compare the differences in prompt-image pairs (\textbf{R2}) and shows the distribution of images (\textbf{R1}, \textbf{R3}, \textbf{R4}).
The system also provides a right panel for users to review the detailed records (\textbf{R1}, \textbf{R4}).
\revision{Users can set the parameters for the embeddings and edges via the control panel on the top. A left panel allows users to create new images.}

\textit{\ivg} (Fig.~\ref{fig:interface}a) is the main view of \promptvis{}.
As described in Section~\ref{sec:graph}, it allows users to navigate the generated images as well as analyze the differences in the prompts and images.
\revision{In addition to the main graph, an embedding mini-map (top left of Fig.~\ref{fig:interface}a) presents the overall node distribution and clusters. The bottom legend allows user to choose the way to present bubbles.}
\ivg focuses on the semantic distribution and relations, and not the details of each step or its temporal order.
To complement this, the \textit{history box} (Fig.~\ref{fig:interface}b) includes the detailed prompting records in chronological order, including all the prompts and the images generated from them.
\revision{The history box also presents detailed modifications in prompts by highlighting the differences of consecutive attempts if they are \textit{similar}, i.e., the similarity is higher than the lower bound $S_{min}$ (see ``text-preprocessing'' in Section~\ref{sec:edge}). The highlights use a consistent color mapping with the glyphs for word change. \textit{Inserted} and \textit{removed} words are in bold style to differentiate from \textit{increase weight} and \textit{decrease weight}.}
\textit{Navigation mini-map} (Fig.~\ref{fig:interface}c) serves as a brief summary of the history records.
\revision{In the mini-map, each prompt is represented by a small dot, the size of which indicates the length of the prompt. The color mapping of the dots is consistent with that of the \ivg, i.e., the temporal order of the prompts.}
\revision{Each pair of \textit{similar} prompts is represented by an arc linking the two corresponding dots. For each dot, the link to prior dots (if exists) with the highest similarity is emphasized with bolder and darker stroke. The line segments on the right of the dots represent different stages of exploration. By default, a step is considered as the beginning of a new stage if the prompt is not \textit{similar} to the previous step. Users can change the division of stages by clicking the gap between two lines to connect them or clicking on a line to divide it.}

\revision{The \textit{control panel} (Fig.~\ref{fig:interface}d) allows user to set the parameters for the visual presentation. The left button ``IVG'' controls whether to show \ivg. The next four sliders set the weight of combining the text and image embeddings, the similarity threshold $S_{min}$, the weight threshold $W_{min}$, and distance threshold to control the number of clusters. Changes in the thresholds will update the \ivg (see the calculation in Section~\ref{sec:edge}). Users can create new sessions and enter prompts through the \textit{creation panel} (Fig.~\ref{fig:interface}e). Currently \promptvis is connected to a Stable Diffusion model (version 1.5) which is open-sourced and fast in generation so that we can easily test the prototype in real-time creation. The other views are updated once new images are generated.}

\begin{figure}[htbp]
    \centering
    \vspace{-0.3cm}
    \includegraphics[width=1\linewidth]{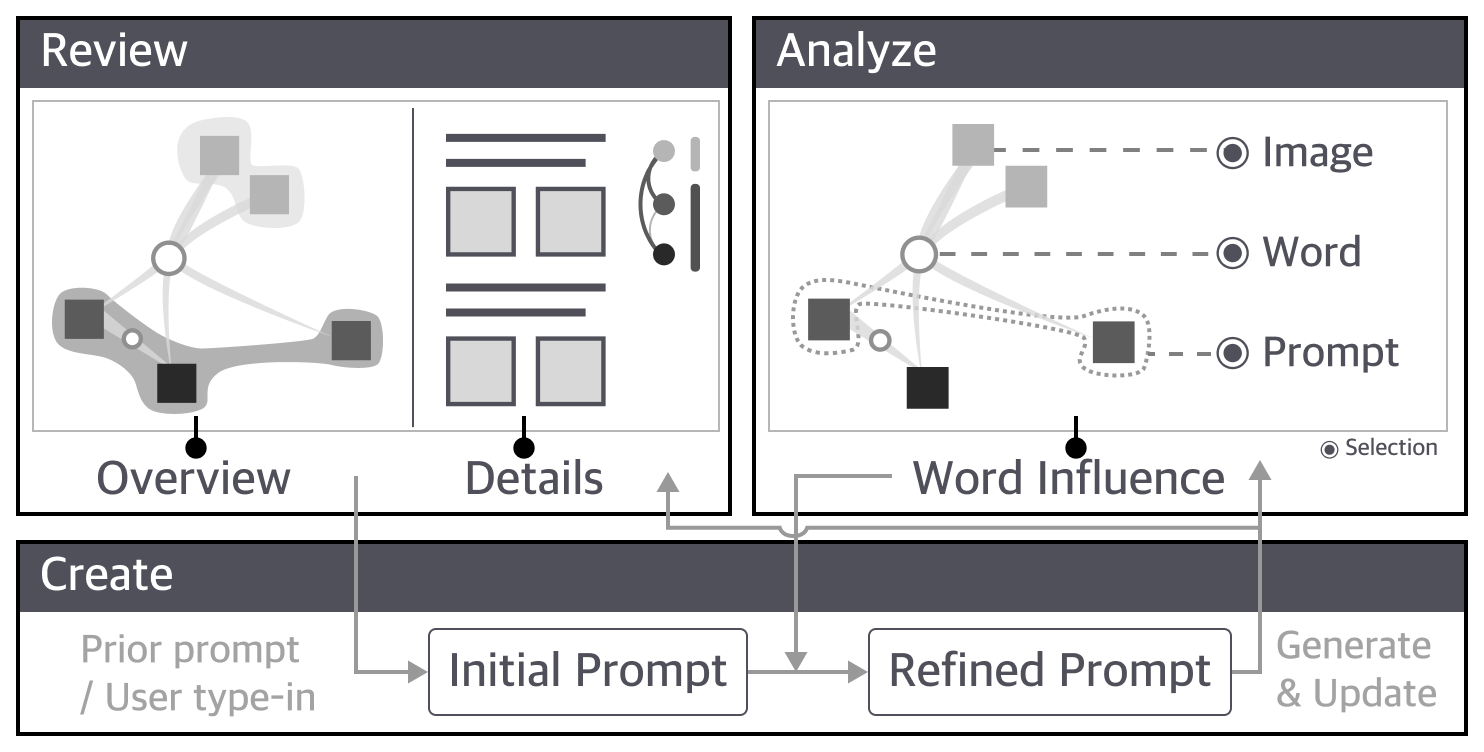}
    \vspace{-0.7cm}
    \caption{\revision{The exploration pipeline of \promptvis. Users can review and analyze previous attempts. They can leverage insights of word influences to refine prompts. The new generation updates the views, allowing further analysis.}}
    \label{fig:explore}
    \vspace{-0.25cm}
\end{figure}

\revision{Fig.~\ref{fig:explore} illustrates the exploration pipeline of \promptvis. Basically, users can review the prompts and images either at a macro level from the \ivg or in a detailed manner from the history box. When observing interesting or desirable images, they can copy the prompt to the input area for next step of generation. Users can also select prompts, images, and words to compare the attempts and analyze the model's behavior. The major insight from such analysis is how the word modification influences the generation, which can be leveraged to decide whether to include certain words in the new attempt and help improve the prompt. More details and examples about the usage of the system can be found in the results of the qualitative evaluation (Section~\ref{sec:interview}).}

\section{\revision{Evaluation}}
\label{sec: user_study}

\revision{The evaluation of \promptvis includes two parts, 
the first one is a quantitative user study to evaluate the system on specific tasks (described in Section~\ref{sec:user_study:ivg}), and the second is a qualitative evaluation with potential users to better understand the system's usability and effectiveness in supporting creative explorations (described in Section~\ref{sec:interview}).}

\vspace{-0.1cm}
\subsection{\revision{Quantitative User Study}}
\label{sec:user_study:ivg}

\subsubsection{Participants and Process}

\revision{We recruited 11 post-graduate students including two females for evaluating the usefulness of \promptvis.}
All participants reported that they have tried generative AI before and graded 4.45/5 on average on their degree of familiarity to text-to-image models.
However, they are less familiar with prompt engineering (3.78/5 on average).

\revision{We aimed to investigate whether \promptvis helps in the review and analysis of prompt history, i.e., \textbf{R1}, \textbf{R2}, and \textbf{R3}, and designed corresponding tasks.
Participants started with training on how to use \promptvis and practiced text-to-image generation through free exploration.
Then, they analyzed three pre-recorded sessions to complete the tasks.
One of the three sessions was manually recorded by the artist we interviewed (Section~\ref{sec: space}), with 16 steps in total. The other two were generated by amateur users that had used the system for open-ended exploration, with 15 steps and 26 steps, respectively.
For each session, users were assigned with three tasks. The first task (\textbf{T1}) was to review the history and identify the exploration stages (\textbf{R1}). The second one (\textbf{T2}) was to compare the prompts between two given image clusters and identify the key words that lead to the variation (\textbf{R2}). The third (\textbf{T3}) was to summarize the model's sensitivity to given words (\textbf{R3}).}

\subsubsection{Results}

\begin{figure}[htbp]
    \centering
    \vspace{-0.3cm}
    \includegraphics[width=1\linewidth]{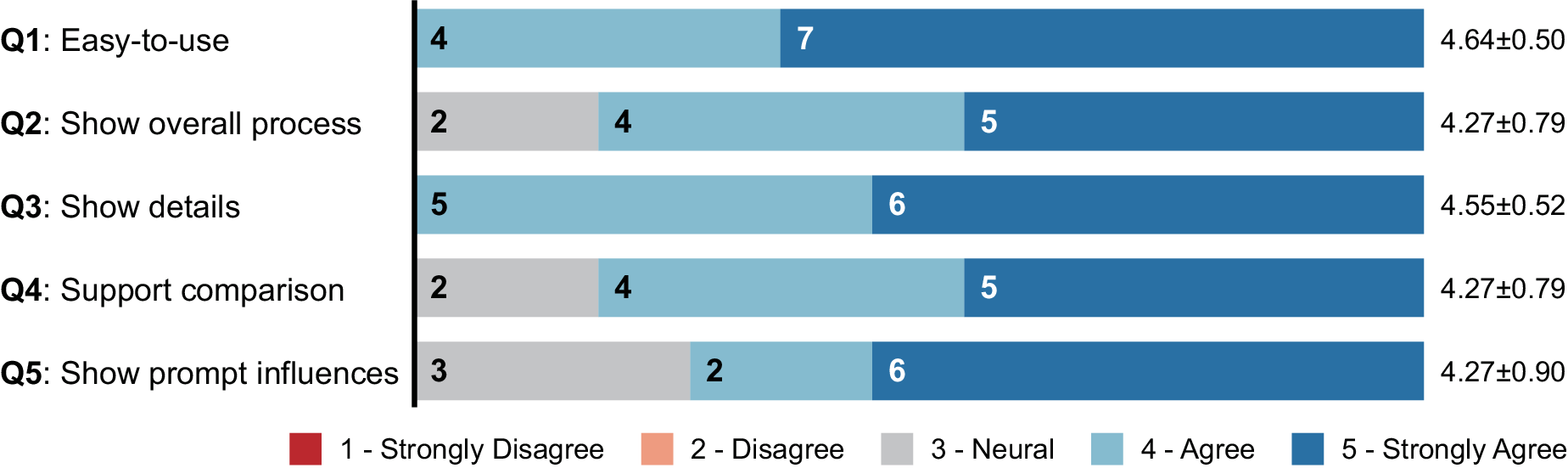}
    \vspace{-0.7cm}
    \caption{Rating for the usefulness of \promptvis in assisting users' analysis of the creative process.}
    \label{fig:user_study}
    \vspace{-0.25cm}
\end{figure}

\revision{For each task, the ground truth is a list or set of descriptions represented as keywords, i.e., a list of themes for \textbf{T1}, a set of words for \textbf{T2}, and a set of keywords describing the word influence for \textbf{T3}. The maximum score is ``5'' if an answer contains all the expected keywords (and in the correct order if the answer is expected to be a list). Otherwise, the score is computed as the proportion of correct keywords out of 5.}
Participants achieved a 82.21\% accuracy to identify the image themes during the creation, which means that participants can easily distinguish the image spaces involved. Besides, most participants reported the differences between clusters correctly (with an accuracy of 96.92\%). However, when asked to identify the influences of certain words, some participants focused on the most salient variation, while overlooked the distinct impact when the word is involved in other context. This leads to a relatively low accuracy (78.12\%). Fig.~\ref{fig:user_study} shows the participants' ratings regarding the usefulness of \promptvis. Participants found the edges especially useful for learning how the changes to words would affect the model's performance, for example, P1 identified magic words which would lead to surprising outcomes.
\revision{The user study demonstrates that \promptvis can help users review the creative process and make sense of the generative model through efficient comparison of prompt-image pairs.}

\vspace{-0.4cm}
\subsection{\revision{Qualitative Study}}
\label{sec:interview}

\subsubsection{Participants and Process}

\revision{To understand how \promptvis can support creative process, we conducted qualitative interviews with both artists and amateur users.
We recruited amateurs as we wanted to investigate whether the system could benefit non-professional users and whether there are differences in the exploration pattern and needs between the two user groups.}

\revision{We recruited 6 users for the interview. Four of them (P1-P4) are university professors who study, teach, and practice visual art. P3 and P4 are the artists that we interviewed during the design stage (Section~\ref{sec: space}). All four participants have more than 20 years of professional art experience, and frequently utilize generative models to explore and pre-produce artistic ideas. P1, P2, and P3 primarily use Midjourney and have also tried Stable Diffusion, DALL-E and GPT4V, while P4 is familiar with Disco Diffusion and Blockade labs Skybox. P5 is a designer with 7 years of art experience  in interaction design, who has used Midjourney before but not very experienced in generative art. P6 is a postgraduate student majoring in computer science who has not received professional art training and only have several attempts at Stable Diffusion. In terms of the familiarity with visualization and visual analytics, P4 had basic knowledge of visualization charts, P1, P2, and P5 were familiar with information visualization but not experienced with visual analytics, while P3 and P6 specialized in this field.}

\revision{The interview began with a 15-minute training on the usage of the system. Participants that had less experience with generative models were given a bit more time (around ten minutes) to do a creative session with the baseline system (a limited version of \promptvis, i.e., only history box and mini-map) so that they could compare the experience and that with \promptvis.
Then, participants had around 20 minutes to iteratively generate images for a topic. Participants can either propose their own topic or choose one from 10 pre-selected topics, which are conceptual themes commonly discussed in abstract art and allow a vast exploration space. Participants were encouraged to think aloud during the process.
After that, participants used the system to recount their creative session within 10 minutes. The last 15 minutes was a semi-structured interview on participants’ experience and feedback.}

\subsubsection{\revision{Data Analysis}}

\revision{All the sessions were conducted through video conferences, which were recorded and transcribed. The prompts and images created in the interviews were automatically recorded by the \promptvis system.
We conducted a thematic analysis~\cite{braun2006using} of the interview data.
We started with theoretical analysis using the requirements as the themes with a special focus on how the requirements were fulfilled (described in Section~\ref{sec:interview}-3).
Then, we conducted inductive analysis of other information relevant to the use of the system and generative AI (discussed in Section~\ref{sec:interview}-4).}

\subsubsection{Results}
\label{sec:interview_results}

\revision{Based on the observation and feedback from the interview, we summarize the instances of the targeted tasks related to the requirements and how \promptvis help users complete them.

\paratitle{R1. Review.} All participants frequently used history box to review their attempts and identify desirable images based on which to refine the prompts. In comparison, \ivg was typically referred to after a certain period of exploration, serving as an overview and provided a new perspective on the images and creative process.

\begin{itemize}[leftmargin=*]
    \item \textbf{Go back to previous attempts.} It is a common that the generated images deviate further from expectations after several steps. In such cases, participants were observed to locate a previous attempt that was relatively satisfactory using the history box. P1 found mini-map particularly useful as \textit{``it illustrates the recursive modifications and helps pinpoint the frequently revisited attempts.''}

    \item \textbf{Help recount creative session.} Participants could quickly recall their previous attempts with \ivg in the recount session, \textit{``especially when there is a large amount of steps''} (P3). P1 and P6 also found the edges useful for grasping the major changes they attempted.
    
    \item \textbf{Provide a new perspective.} P1 and P4 commented that \ivg provided them with a different perspective, which is helpful for understanding and navigation. \textit{``The graph enables me to ignore the prompts and look at the images on their own merits''} (P4). Similarly, P1 noticed some images, though not aligned with the initial intention, looked good in other aspects.
    
\end{itemize}

\paratitle{R2. Compare.} During creation, P1, P5, and P6 engaged in comparing the prompts and images through \ivg. P2, P3, and P4, however, focused on the history box, observing whether the outputs align with their intentions.

\begin{itemize}[leftmargin=*]
    \item \textbf{Observe result of prompt change.} While it is easy to compare consecutive attempts in the history box, P1 found it more intuitive to observe the change on \ivg, as \textit{``it indicates the distance between images and different levels of impacts of the changed words.''}

    \item \textbf{Search for similar images.} Upon obtaining a satisfactory image, P5 explored its neighbors in \ivg and identified two other similar images. He then compared the prompts of the three images, which were quite different, and selected the common phrases for the new attempt.

    \item \textbf{Observe prompt stability.} All participants agreed that the bubbles indicating images generated by the same prompt increased their awareness of prompt stability. \textit{``The images generated by one prompt can be quite different. I believe the visualization helps to identify more stable prompts'' (P2).}
\end{itemize}

\paratitle{R3. Model behavior.} We observed that \promptvis help participants improve the knowledge about the general model behavior, e.g.,
the model is not good at creating things that do not exist,
as they experimented with different prompts, but \textit{``the more you explore the same idea, the more muddy it gets. It's like casting a fishing line, and if you throw it in different spaces, you get different versions''} (P4). \ivg could reduce such confusion, providing additional insights into the macro model characteristics.

\begin{itemize}[leftmargin=*]
    \item \textbf{Distinguish influence of certain phrase.} It could be tricky to distinguish the roles of different stylistic and descriptive terms when they are mixed in one prompt. P5 went through trials and errors with different combination of phrases in the baseline session but could not identify any clear rule. When recounting the session with \ivg, he realized that ``light red'' somewhat conflicted with ``Chinese painting,'' adding modern elements to the outputs, which were expected to be in the traditional style.

    \item \textbf{Reasoning causes of unsatisfactory images.} During the creative session with \promptvis, P5 got a bit stuck and could not further improve the outputs. By examining the stage bubble on \ivg, P5 identified the edge, i.e., inserting ``album cover,'' which contributed to the group of unsatisfactory images. This observation helped P5 remove the phrase in following attempts, leading to better results.

    \item \textbf{Recall insights to control generation.} Sometimes, participants might forget the knowledge accumulated during the exploration.
    The history box and \ivg served as a reminder and assisted P3 in recalling the knowledge gained to avoid repetitive failures.
    For example, once P3 experimented with apple in the style of ``ruthko'' and was not satisfied. By reviewing the previous session, which was to generate a cartoon pig, and observing the edges of replacing less popular styles with ``Monet,'' P3 recalled that the model performed better on the style of ``Monet'' and decided to generate an apple in this style.
\end{itemize}

\paratitle{R4. Plan.} Some observations discussed so far already showcased how participants designed new prompts with the aid of \promptvis. Here we summarize some typical exploration patterns of the participants and demonstrate how the system facilitates the planning.

\begin{itemize}[leftmargin=*]
    \item \textbf{Improve prompts towards a clear target.} P1, P2, and P5 had a target image in mind before starting, which is a common practice in AI-assisted design and pre-production. P2 appreciated the rationality and effectiveness of \ivg in assisting prompt engineering, \textit{``the tool makes sense to me as it can help me understand how to improve my prompts.''}

    \item \textbf{Adjust the target according to outputs.} In non-industrial settings, there is usually more flexibility in deciding the final representation of an art work. While started with a certain goal, P3 adapted the generated scene to model's capability, e.g., involving more characters in the story when the model failed to generate images with exact one character. \promptvis helped P3 make adjustments based on knowledge of model behavior (as described in R3).

    \item \textbf{Explore realizations of abstract idea.} P4 aimed to explore a film idea, which was more conceptual and open. Throughout the exploration there were quite a few inspiring images that served as starting point of new branches and variations of the idea, which the participant frequently went back to through history box to start a new series of attempts.

    \item \textbf{Help build exploration mental map.} P6 found \ivg effective in revealing the unexplored space and helpful to construct a mental map of the creative space. \textit{``When creating with the baseline system, I often focused on the most recent steps and was unwilling to branch out.''} \ivg, however, reminded P6 of the previous attempts, motivating and guiding him to combine the knowledge learned in both stages and identifying unexplored space that might be promising. \textit{``The graph helps me adjust the combination, I could imagine where the desired results are in the embedding space and fine-tune prompts accordingly.''} %
\end{itemize}}

\subsubsection{\revision{Discussion}}

\revision{On the whole, all participants agreed that history box is helpful and critical to the creative process. While P1, P5, and P6 leveraged \ivg for real-time planning, P2, P3, and P4 thought \ivg is more useful for reviewing previous attempts. %

\paratitle{Attention and interest.} Participants' preferences on the amount of information shown in \ivg vary when they have different tasks. P1, P2, and P3 suggested that since the capacity of human attention is limited, during the creative process, it would be distracting if we present too many nodes and edges on the graph without distinguishing the levels of emphasis. P1 and P2 proposed using the size of image to encode levels of user interest. In contrast, when engaged in review and analysis, P4 found it more effective to show more images, especially those similar to the desired ones. P5 and P6 demanded more textual information, e.g., common phrases in prompts of focused images, to aid them in refining prompts.
Currently the \promptvis  is mainly designed to support review and recount. We will include these feedback in the future work for guidance and recommendation.

\paratitle{Capture complete context.} With the rapid development of generative models and tools, prompt engineering is not the only way to control the generation. For example, P1 and P3 uses hand-drawn sketches or their own artwork as image seeds to specify the desired object and style. Though \promptvis currently focuses on prompt-image pairs, it is important to take other context, e.g., seed image, parameter, and image editing, into consideration. Besides, P2 envisioned the capability to comparing different models, and it would be interesting to capture and integrate the explorations across tools.

\paratitle{Organization and curation.} Currently \promptvis allows users to organize their attempts into different sessions. However, P2 wished more advanced and flexible organization, such as tagging the images and arranging them along a story line (if the goal is to explore ideas around a film). We also observed that P4 saved and annotated inspiring outcomes in a document as externalization of the ideas, so that he could reflect on what resonates with the original idea later. All the artists expressed their willingness and needs to curate the exploration history, e.g., rating and pinning the generated images, taking notes of the attempts, etc. Such organization and curation could form part of context in creative provenance and be leveraged to infer user intention and preference.

\paratitle{Accurate understanding of user preference.} All the participants with professional art training attempted to accurately control the outputs to realize their goals. They either had a clear picture in mind before the generation, or had accurate senses of the desired features, e.g., the composition, environment, and emotion, even though the process can be exploratory. For the latter case, \textit{``translating internal senses and feelings into prompts that the model can understand becomes even more important and challenging''} (P3). The senses and preferences are a reflection of the artist's style and inspiration. P3 expressed the concerns with the ``creativity'' with current generative models, \textit{``simply combining many styles and elements together might create something that looks new, but it is like to go from 1 to 99, instead of creating something original.''} Dataset-based or LLM-based recommendations have been proven to generate appealing images favored by public users, but the artists hope the model to truly understand their personal styles and artistic tastes. It is a promising direction to understand user preference and make recommendations based on exploration provenance.}

\section{\revision{Conclusion and Future Work}}
\label{sec: conclude}

This work proposes \ivg to help artists understand the influences of prompt modifications during text-to-image generation. It is part of the \promptvis, a visual analytics system for users to review and understand the prompting history for a more effective creative process. 
The \ivg models the differences in prompts and their impacts as edges between image nodes that are projected according to their text and image similarity. 
Thus, users can observe the semantic distribution as well as analyze the effects of prompt modifications.
\promptvis allows users to directly interact with a generative image model, a time-oriented view for prompting history, and several additional features to support the creative process.
Both a quantitative and a qualitative user study were conducted to evaluate the effectiveness of \ivg and \promptvis. Participant highly rated the usability of both, and the qualitative results revealed how the features in \ivg and \promptvis help user better completing the targeted tasks.

\revision{Generative art, which involves both human and AI models to achieve creative goals, has presented challenges and opportunities for visualization research.}
\promptvis is an initial step towards understanding and utilizing individual creation history.
\revision{Based on the results and feedback we received, future work will focus on the following aspects:
\begin{enumerate}
\item Improvement of the methods, layout, and encoding of \ivg to enhance readability and usability for better real-time support.
\item Complete provenance capture and support for a wider range of user types in more realistic settings.
\item Personalized recommendation for artists based on exploration provenance and user preference.
\end{enumerate}}

\section*{Acknowledgments}
This work is supported by NSFC No.~62272012.

\bibliographystyle{IEEEtran}
\bibliography{template}

\vspace{-0.8cm}

\begin{IEEEbiography}[{\includegraphics[width=1in,height=1.25in,clip,keepaspectratio]{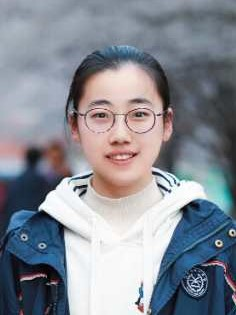}}]{Yuhan Guo}
is a PhD student at the School of Intelligence Science and Technology, Peking University. She received a B.S. degree in intelligence science and technology from Peking University in 2023. Her research interests include text visualization and visualization for humanities. Her recent research focuses on visualization of provenance data and visual analytics for sensemaking tasks through human-AI collaboration.
\end{IEEEbiography}

\vspace{-0.7cm}

\begin{IEEEbiography}[{\includegraphics[width=1in,height=1.25in,clip,keepaspectratio]{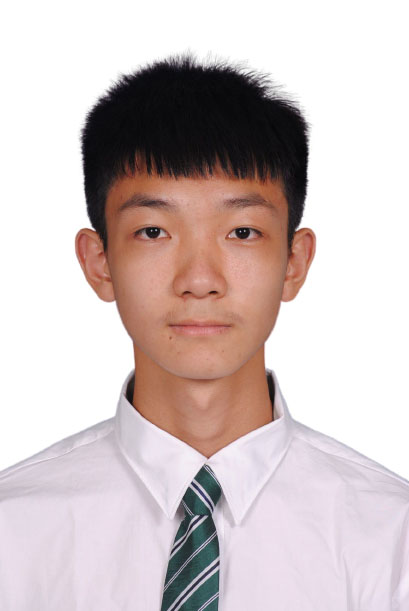}}]{Hanning Shao}
is now a Ph.D. student at the School of Intelligence Science and Technology, Peking University.
He received a B.S. degree in computer science from Peking University in 2021. His research interests include scientific visualization.
\end{IEEEbiography}

\vspace{-0.7cm}

\begin{IEEEbiography}[{\includegraphics[width=1in,height=1.25in,clip,keepaspectratio]{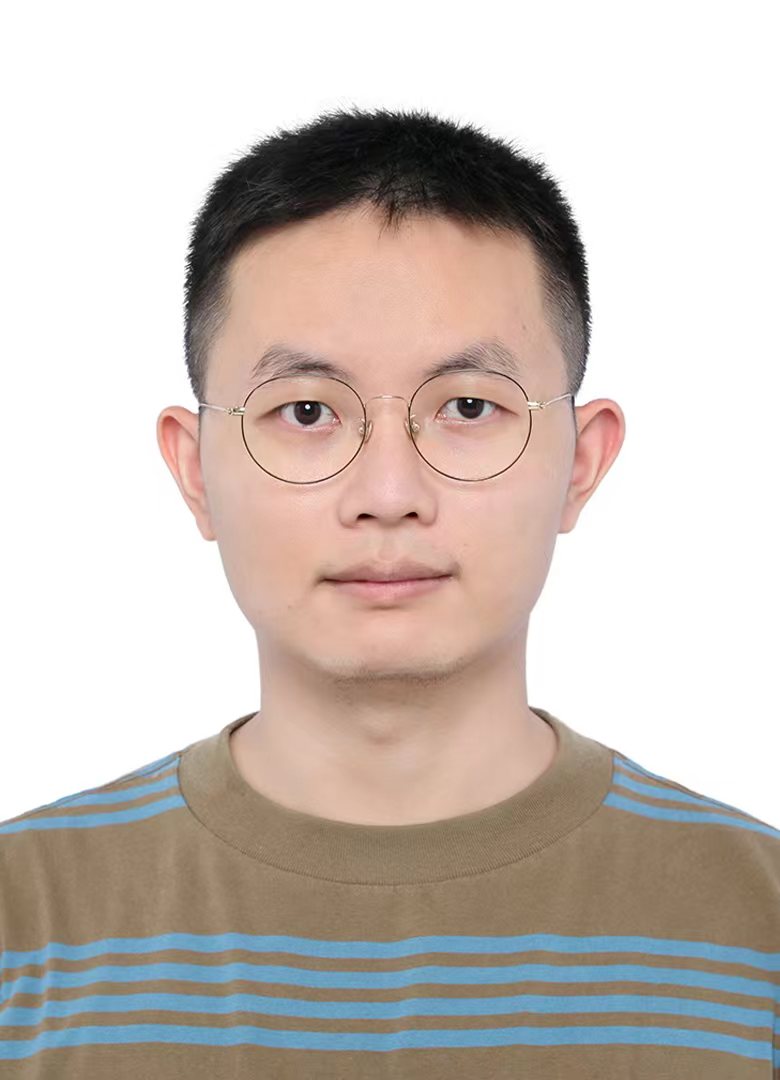}}]{Can Liu}
received a B.S. degree in computer science and a B.E. degree in economics from Peking University in 2018, and received a Ph.D. Degree at the School of Intelligence Science and Technology, Peking University in 2023.
His research interests lie in the field of deep learning-driven visualization, especially intelligent interaction for visualization.
\end{IEEEbiography}

\vspace{-0.7cm}

\begin{IEEEbiography}[{\includegraphics[width=1in,height=1.25in,clip,keepaspectratio]{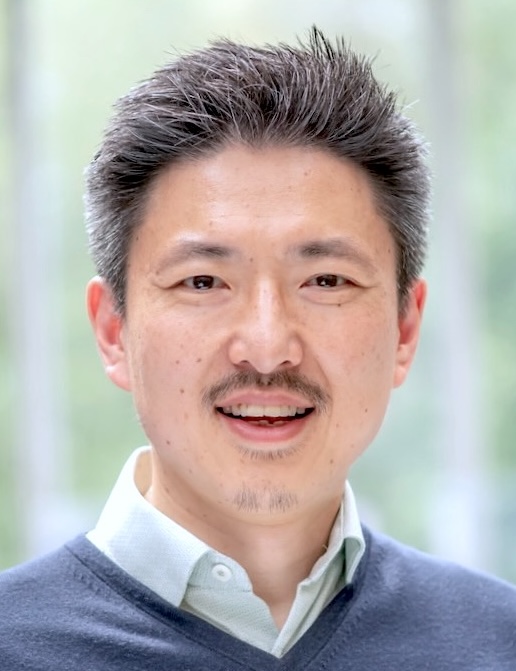}}]{Kai Xu} is an Associate Professor in the School of Computer Science at the University of Nottingham in UK. He is the co-director of School’s Visualization Research Group. His main research interest is Data Science, particularly Data Visualization. His recent work focuses on designing interactive visual interfaces for human-AI teaming. He received his BEng in Computer Engineering from Shanghai Jiaotong University in 1999 and later PhD in Computer Science from Univesrity of Queensland in Australia in 2004.
\end{IEEEbiography}

\vspace{-0.7cm}

\begin{IEEEbiography}[{\includegraphics[width=1in,height=1.25in,clip,keepaspectratio]{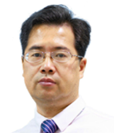}}]{Xiaoru Yuan} received a B.S. degree in computer science and a B.A. degree in law from Peking University in 1997 and 1998, respectively. In 2005 and 2006, he received an MS degree in computer engineering and a Ph.D. degree in computer science from the University of Minnesota.
He is now a professor at Peking University in the Laboratory of Machine Perception (MOE). His primary research interests lie in scientific visualization, information visualization, and visual analytics, emphasizing large data visualization, high dimensional data visualization, graph visualization, and novel visualization user interface. He is a senior member of the IEEE.
\end{IEEEbiography}

\vfill

\end{document}